\documentclass[grl]{AGUTeX}

\usepackage[dvips]{graphicx}

\authorrunninghead{LUGAZ AND FARRUGIA}

\titlerunninghead{A NEW CLASS OF COMPLEX EJECTA}

\authoraddr{Corresponding author: N. Lugaz,
Department of Physics and Space Science Center, University of
New Hampshire, Morse Hall, 8 College Rd, Durham, NH 03824, USA.
(noe.lugaz@unh.edu)}

\begin{document}

%% ------------------------------------------------------------------------ %%
%
%  TITLE
%
%% ------------------------------------------------------------------------ %%

\title{A New Class of Complex Ejecta Resulting From the Interaction of Two Coronal Mass Ejections With Different Orientations and Its Expected Geo-Effectiveness}

%% ------------------------------------------------------------------------ %%
%
%  AUTHORS AND AFFILIATIONS
%
%% ------------------------------------------------------------------------ %%

 \authors{N Lugaz \altaffilmark{1}
 and C.~J. Farrugia\altaffilmark{1}}

\altaffiltext{1}{Department of Physics and Space Science Center, University of
New Hampshire, Durham, NH, USA.}

\begin{abstract}
A significant portion of transients measured by spacecraft at 1~AU does not show the well-defined properties of magnetic clouds (MCs). %Among these non-MC ejecta, some are thought to be associated with the interaction of successive coronal mass ejections (CMEs). %A particular focus has been on recognizing and understanding multiple-MC events, where two MCs cross the spacecraft path one after the other with signs of interaction between them. 
Here, we propose a new class of complex, non-MC ejecta resulting from the interaction of two CMEs with different orientation, which differ from the previously studied multiple-MC event. At 1~AU, they are associated with a smooth rotation of the magnetic field vector over an extended duration and do not show clear signs of interaction. We determine the characteristics of such events based on a numerical simulation and identify and analyze a potential case in the long-duration CME measured {\it in situ} in 2001 March 19--22. Such events may result in intense, long-duration geo-magnetic storms, with sawtooth events, and may sometimes be misidentified as isolated CMEs. 
\end{abstract}

%% ------------------------------------------------------------------------ %%
%
%  BEGIN ARTICLE
%
%% ------------------------------------------------------------------------ %%

%
% \end{article} must follow the references section, before the figures
%  and tables.

\begin{article}

%% ------------------------------------------------------------------------ %%
%
%  TEXT
%
%% ------------------------------------------------------------------------ %%

\section{Introduction}
Coronal mass ejections (CMEs) are the major driver of intense geo-magnetic activity \citep[]{Richardson:2001} and have been studied extensively for the past 40 years. CMEs are observed remotely by coronagraphs and heliospheric imagers and measured {\it in situ} by spacecraft such as ACE and {\it Wind}. With the launch of SOHO and STEREO, the availability of white-light imagers with wide fields-of-view has made it possible to associate eruptions observed in the corona to CMEs measured {\it in situ} at 1~AU \citep[see for example the list by][]{Richardson:2010}. CMEs measured {\it in-situ} may be divided into three broad categories: magnetic clouds (MCs), non-MC isolated ejecta, and complex ejecta \citep[similar to the categories of][]{Zurbuchen:2006}. MCs have well-defined properties \citep[see][]{Burlaga:1981}. Non-MC isolated ejecta typically have some but not all the properties of MCs, and may be sometimes referred to as MC-like ejecta \citep[]{Lepping:2005}. They may correspond to a distorted CME or to the crossing through the ``leg'' of a CME. Lists of MCs and MC-like ejecta measured at 1~AU by the Wind and ACE spacecraft are maintained \citep[]{Lepping:2005,Jian:2006,Richardson:2010}.

Complex ejecta result from the interaction of successive CMEs \citep[]{Burlaga:1987}. Some consist of many individual eruptions and it is impossible to relate {\it in situ} measurements to coronagraphic observations of CMEs \citep[]{Burlaga:2002}. Others are made up of two clearly distinct MCs separated by an interaction region \citep[multiple-MC events, see][]{Wang:2003, Lugaz:2005b}. Complex ejecta tend to have long duration and may drive the magnetosphere for an extended period. \citet{Xie:2006}, for example, studied long (3 days or more) and intense (peak Dst $\le -100$ nT) geomagnetic storms and found that 24 out of  37 such storms were associated with multiple CMEs. 

While a typical CME passes over Earth in $\sim 20$ hours, some events have duration well in excess of 30 hours \citep[]{Marubashi:2007}. It is possible that some of these long-duration events, believed to be associated with a single, isolated CME are in fact the results of the interaction of two CMEs, a possibility raised for the 2005 May 15 CME by \citet{Dasso:2009}. 

Here, we identify a new type of complex ejecta due to the interaction of two CMEs, which results in a long-duration event with a smooth rotation of the magnetic field vector. In section \ref{simu}, we present the result at 1~AU of two simulations, one of an isolated CME and one of two interacting CMEs, and we discuss the expected geo-effectiveness of such events. In section \ref{geo}, we present  measurements of the 2001 March 19--22 period, which may be associated with the interaction of two CMEs in a way similar to that of the simulation. We discuss our findings and conclude in section \ref{conclusion}.

\section{Simulated Magnetic Cloud and Complex Ejecta}\label{simu}
\subsection{Simulation Set-up}
The simulation set-up is in nearly identical to that of \citet{Lugaz:2013b} for their Case C (two CMEs with orientation $90^\circ$ apart). We summarize the important details here as well as one difference with this previous simulation, and refer the interested reader to this paper for further information. We use the Space Weather Modeling Framework \citep[SWMF,][]{Toth:2012} to perform the simulations. The simulation domain is a Cartesian box centered at the Sun and extending to $\pm 220~R_\odot$ in all three directions. The domain is resolved with a maximum of 34 million cells ranging in size from 0.01 to 4~$R_\odot$ after adaptive mesh refinement (AMR). Along the Sun-Earth line, cell size of 0.05~$R_\odot$ is maintained up to 0.4~AU (0.1~$R_\odot$ thereafter).

 \begin{figure*}[t]
\centering
 \includegraphics[width=7.3cm]{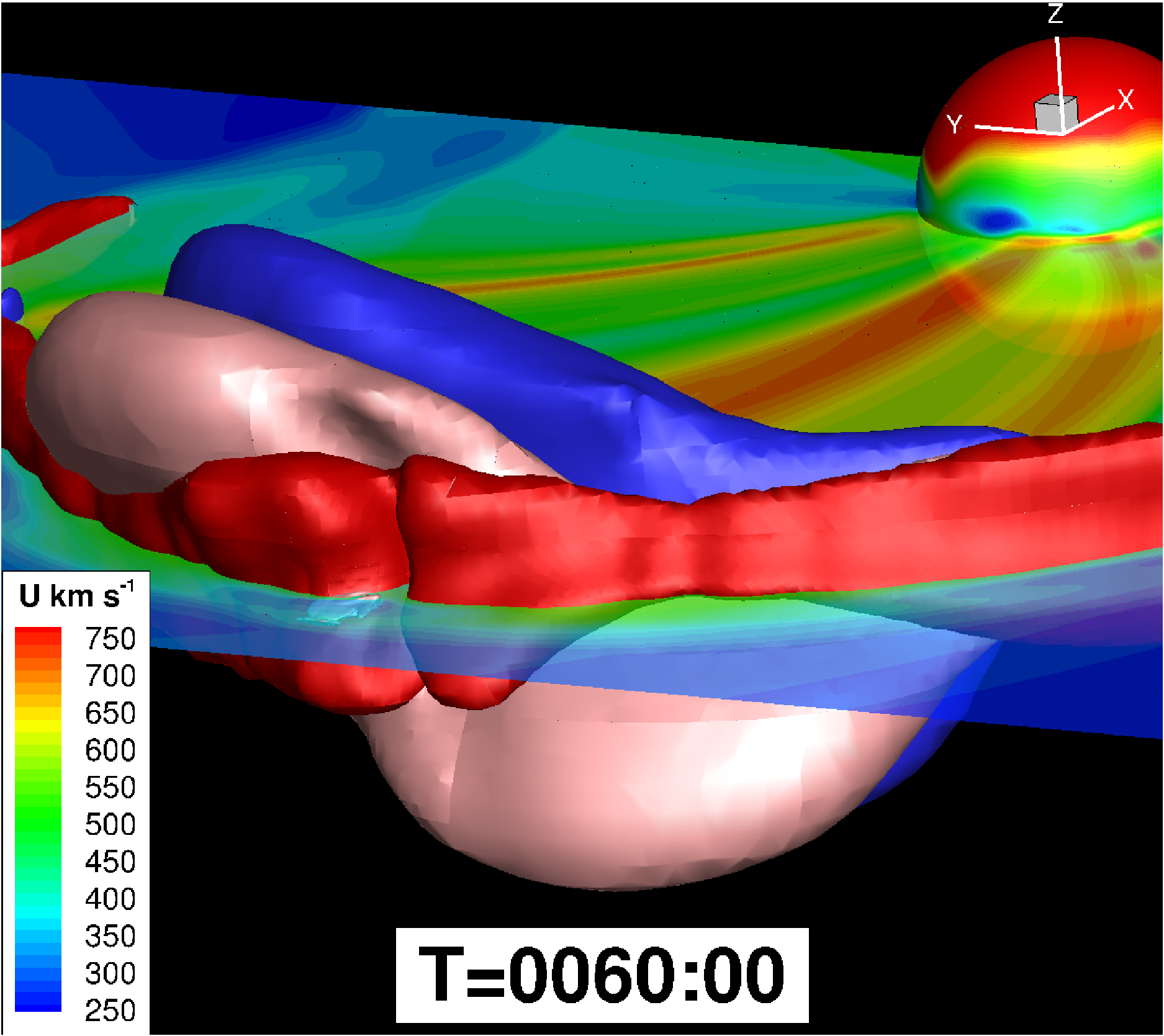}
 \includegraphics[width=7.3cm]{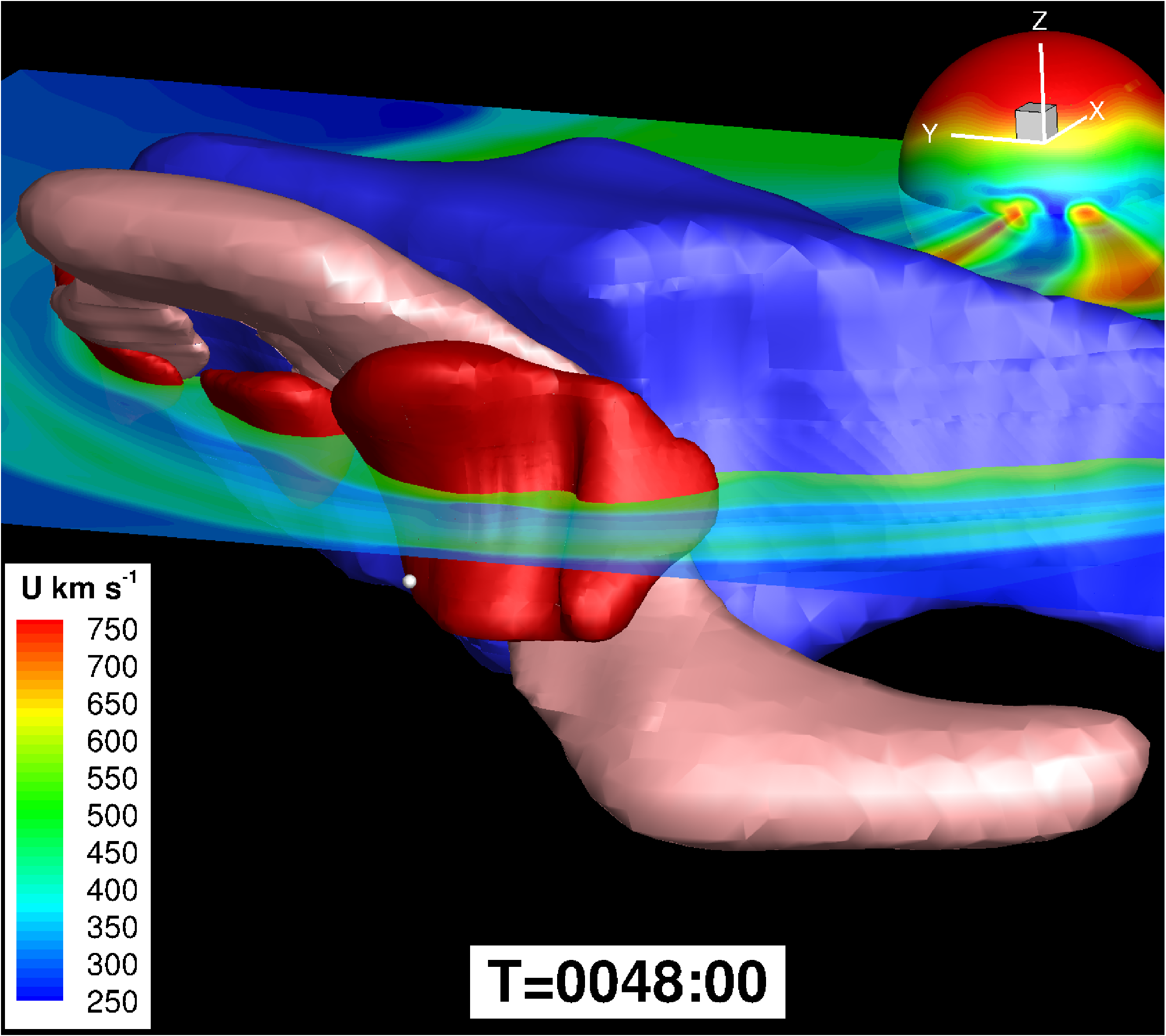}
 \caption{{\it Left}: Simulation result after 60 hours corresponding to one isolated CME. The pink isosurface corresponds to values of $B_y = 20$~nT; the red and blue isosurfaces correspond to values of $B_z = \pm 20$~nT, respectively. The sphere of radius 20~$R_\odot$ centered at the Sun's position and the cut in the ecliptic plane are color-coded with the velocity. {\it Right}: Simulation result after 48 hours corresponding to the two interacting CMEs with the same conventions as the image of the left panel.}
 \end{figure*}

We use the solar wind model of \citet{Holst:2010} where Alfv{\'e}n waves drive the solar wind. To set-up the solar magnetic field, we use a non-tilted dipole with an octopole component, which yield a maximum magnetic field strength of 5.5~G at the solar poles with a polarity corresponding to that of solar cycle 24. To initiate the CMEs, we insert right-handed flux ropes using the model of \citet{Gibson:1998} (GL) in a state of force imbalance onto the steady-state solar corona. The parameters of the GL flux rope for the two CMEs are the same as for the Case C from \citet{Lugaz:2013b}, with the only difference being the time delay between the two CMEs, which is 15 hours instead of 7 hours. The first CME  (CME1) has a low inclination and an eastward axial field: a NES cloud according to the categorization of \citet{Bothmer:1998}, the second CME (CME2) is highly inclined with a southward axial field: a ESW cloud. For comparison purposes, we also perform the simulation of an isolated event  by simply propagating the first CME all the way to 1~AU without the second eruption.

\subsection{Simulated Magnetic Cloud at 1~AU}
Three-dimensional results of the simulation of the isolated CME are shown in the left panel of Figure~1 with isosurfaces of magnetic field equals to 20~nT highlighting the CME. The magnetic ejecta has a reverse-S shape, characteristic of the GL model \citep[see,][]{Gibson:1998,Manchester:2004a}. Synthetic spacecraft measurements at 1~AU are shown in the left panel of Figure~2. It corresponds to a moderately fast CME with a transit time of about 63 hours for the shock and 70 hours for the magnetic ejecta. The CME has a speed at 1~AU of about 540~km~s$^{-1}$ and is characterized by a NES rotation. The sheath duration is about 7 hours and the magnetic ejecta lasts for about 23 hours, corresponding to a width of $\sim$0.28~AU.

\subsection{Simulated Complex Ejecta at 1~AU}
Next, we discuss the results of the simulation for the two interacting CMEs. The timing of the interaction goes as follows: the shock driven by CME2 reaches the back of CME1 at 18.5 hours, its center at 22.5 hours and the back of the sheath at 24 hours. By 29 hours, at 0.47~AU, the two shocks have merged. The results after 48 hours, as the complex ejecta is close to 1~AU, are shown in the right panel of Figure~1. At the back of CME1, there is an extended period of southward magnetic field, as is clear from the large blue isosurface in Figure~1.

Synthetic spacecraft measurements a 1~AU are shown in the right panel of Figure~2. A single fast-mode shock at 54 hours precedes the complex ejecta starting at 63 hours. The complex ejecta is characterized by a relatively short period of northward $B_z$ lasting about 5 hours, and an extended ``tail'' of southward $B_z$ for about 28 hours. The east-west component of the magnetic field, $B_y$ is close to zero for the last 24 hours of the event, after an initial period of eastward magnetic field. Throughout the complex ejecta, the velocity profile is decreasing from about 540 to 450~km~s$^{-1}$.  

The main differences between an isolated MC and the complex ejecta resulting from the interaction of two CMEs are: (i) shorter transit time of the complex ejecta as compared to the isolated CME, (ii) short duration of the first CME (here about 8 hours vs. 28 hours for the isolated one), (iii) hotter and somewhat denser sheath region preceding the complex ejecta.

 \begin{figure*}[t]
\centering
\includegraphics[width=6.5cm]{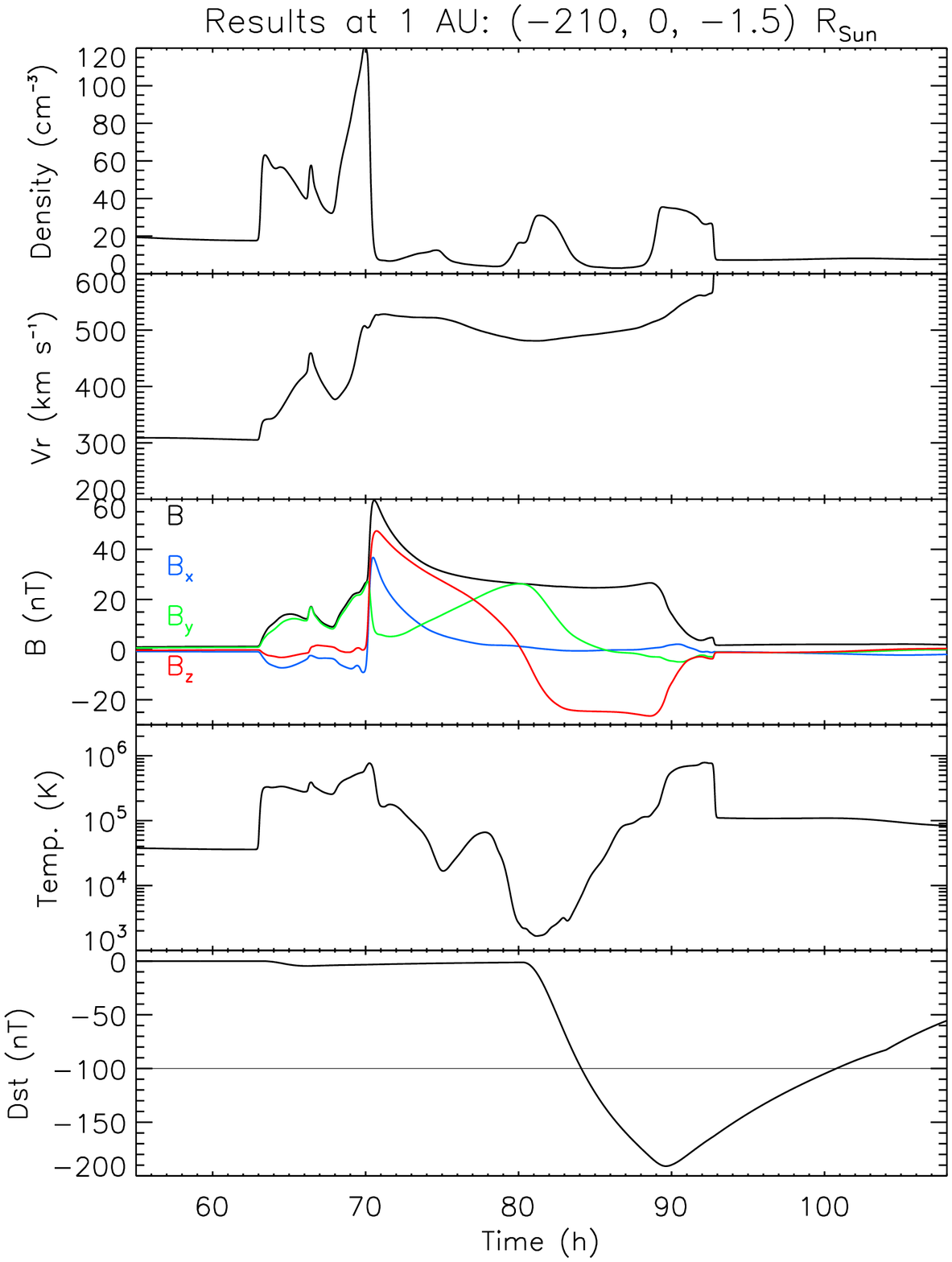}
\hspace{0.7cm}
 \includegraphics[width=6.5cm]{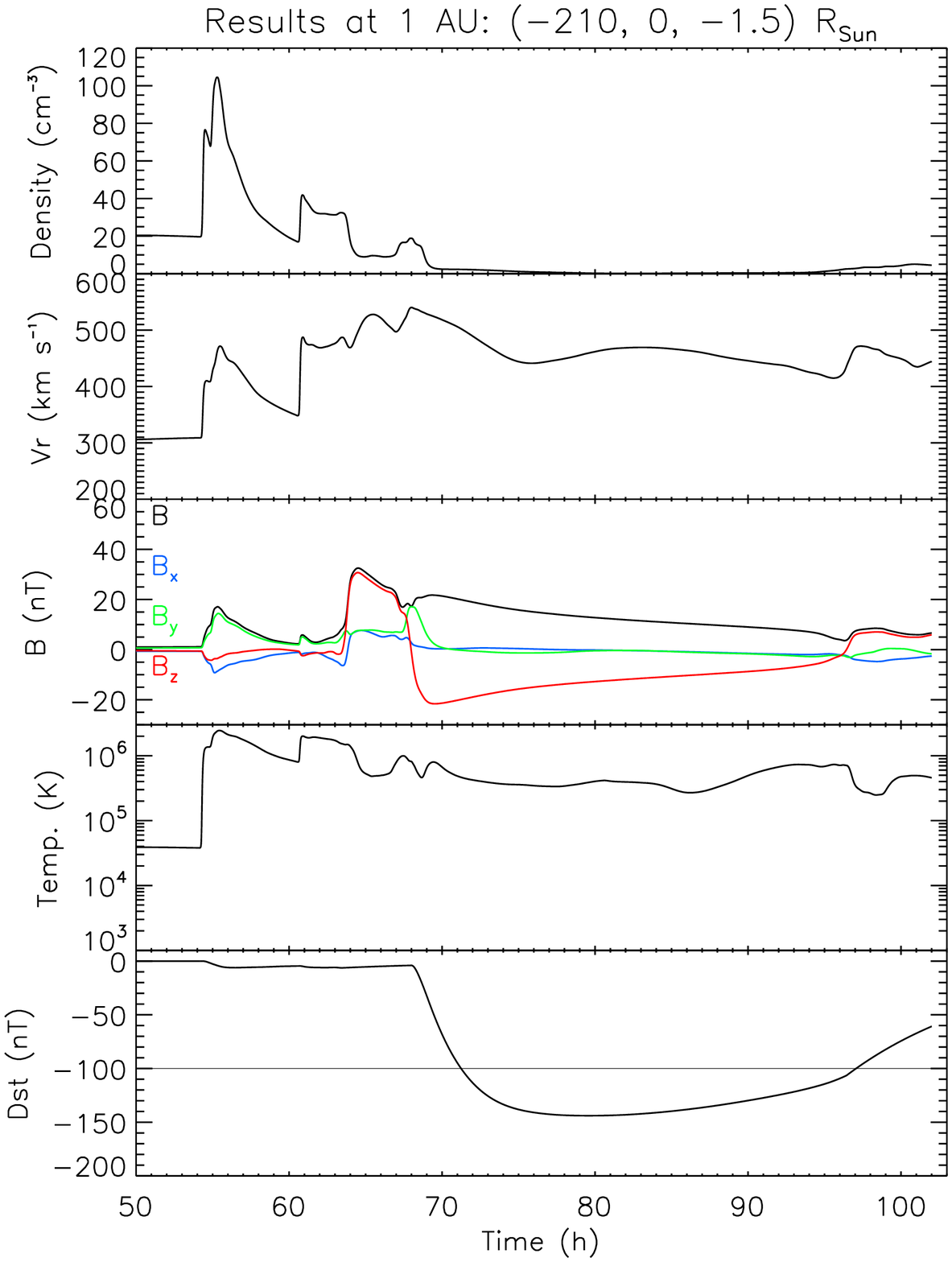}
 \caption{{\it Left}: Simulation result at 1~AU of an isolated CME and modeled Dst index. {\it Right}: Simulation result at 1~AU of a complex ejecta corresponding to the two interacting CMEs described in the text. The panels from top to bottom show the density, radial velocity, magnetic field, temperature and derived Dst with the same scales (except that the complex ejecta is shifted forward by 5 hours).}
 \end{figure*}

\section{Geo-effective Potential of Complex Ejecta}\label{geo}

\subsection{Simulated Events}

We estimate the Dst corrected from the contribution of the ram pressure, $Dst^*$, from the simulated {\it in situ} measurements using a modified version of the \citet{Burton:1975} relation: $ \frac{d}{dt} Dst^* = Q(t) - Dst^*/\tau$, with $Q =  -1.22 \times 10^{-3} \left(VB_z - 0.49\right)$ if $VB_z > 0.49$ mV~m$^{-1}$, and $Q = 0$ otherwise, and with $\tau = 8.64 \times 10^3 \exp{\left(9.74/(4.69+VB_z)\right)}$ if $B_z < 0$~nT and $\tau = 3.57 \times 10^4$ otherwise.

The results are shown in the bottom panels of Figure~2. Both the isolated CME and the complex ejecta would have resulted in an intense geo-magnetic storm with a peak Dst below $-100$~nT. The isolated CME results in a relatively typical intense geo-magnetic storm with a main phase lasting about 9 hours, a peak Dst of $-191$~nT following by a recovery phase lasting more than 1 day. The Dst is below $-100$~nT for about 16.5~hours. %84.11->100.77
The complex ejecta, on the other hand, results in a weaker storm with a peak Dst of $-144$~nT. The main phase lasts about 12 hours and the recovery phase lasting about 1.5 days. The Dst is below $-100$~nT for about 26~hours or about 55\% longer than for the isolated CME. %71.244->97.06
Note also that the southward magnetic field in the sheath region results in a very small negative Dst for the isolated and interacting CMEs around time t = 65 and 55 hours, respectively. 

The larger Dst in the isolated CME is due to a combination of factors: the minimum southward $B_z$ is slightly larger ($-26.5$ vs $-21.6$~nT), it occurs at the end of the southward magnetic field period instead of at its beginning, and the velocity in the complex ejecta decreases faster than in the isolated CME (resulting in a smaller dawn-to-dusk electric field). However, the longer period below $-100$~nT predicted for this complex ejecta may result in an intensification of the geo-magnetic response, which cannot be captured by our simple model to evaluate the Dst index.

\subsection{Real Event}

To confirm that this type of complex event indeed exist, we identify potential examples in {\it in situ} data at 1~AU.  Note that \citet{Dasso:2009} discuss a complex event which may be the result of the interaction of two successive CMEs in 2005 May 15. Here, we start from the list of 17 events from \citet{Marubashi:2007} and that from \citet{Xie:2006}. The complex event from 2001 March 19--22 is one of  the best examples of a potential long-duration complex ejecta resulting from multiple CMEs (another candidate, not discussed here, occurred on 2004 April 3--5). 
%%%%%%
Figure 3 shows the {\it in situ} measurements including the clock angle of the IMF, $\theta$, and the merging electric field, $E_{KL} = V \sqrt{B_y^2+B_z^2} \sin^2(\theta/2)$ following \citet{Kan:1979}. The magnetic ejecta lasts for about 57~hours (between the two vertical red lines). It is preceded by a single fast-mode shock at 11:30UT on March 19 (marked by the first vertical line). The velocity profile through the CME is similar to that of an isolated expanding event: monotonously decreasing with a center value of 400~km~$^{-1}$ and an expansion speed of 100~km~s$^{-1}$. The magnetic field strength is very smooth and reaches a maximum of $\sim 22$~nT and the plasma $\beta$ is below 0.1 throughout the structure. There are however some indications of an origin from two structures with an interface around 18-20UT on March 20: (i) the fluctuations in the magnetic field vectors occur in all three components from the start of the event to 20 UT on March 20; thereafter only the $B_x$ and $B_z$ component fluctuates but the $B_y$ component is smooth; (ii) the magnetic clock angle varies (first decrease to 180$^\circ$ then increase back to $90^\circ$) during the first part and is steady (eastward directed) afterwards, and, (iii) the merging electric field implies strong forcing of the magnetosphere during the first part (large values $\sim 5$ mV/m) and decreases monotonically thereafter. Also, small-scale structures (identified as slow shocks) are present near the peak of the magnetic-field strength (March 20 around 18-20 UT).
%The magnetic field vector is complex and may indicate two separate ejecta. A multiple-MC event, in additional to a single speed profile, is characterized by an interaction region between the two MCs with a lower magnetic field strength, higher proton density, temperature and plasma $\beta$ \citep[]{Wang:2003}. There is no indication of such an interaction region here; in particular, the temperature and plasma $\beta$ are low throughout for more than 2~days. There is, however, an increase in the proton density around 12UT on March 20. 

%%%%%%
 \begin{figure}[t]
\centering
\includegraphics[width=8.3cm]{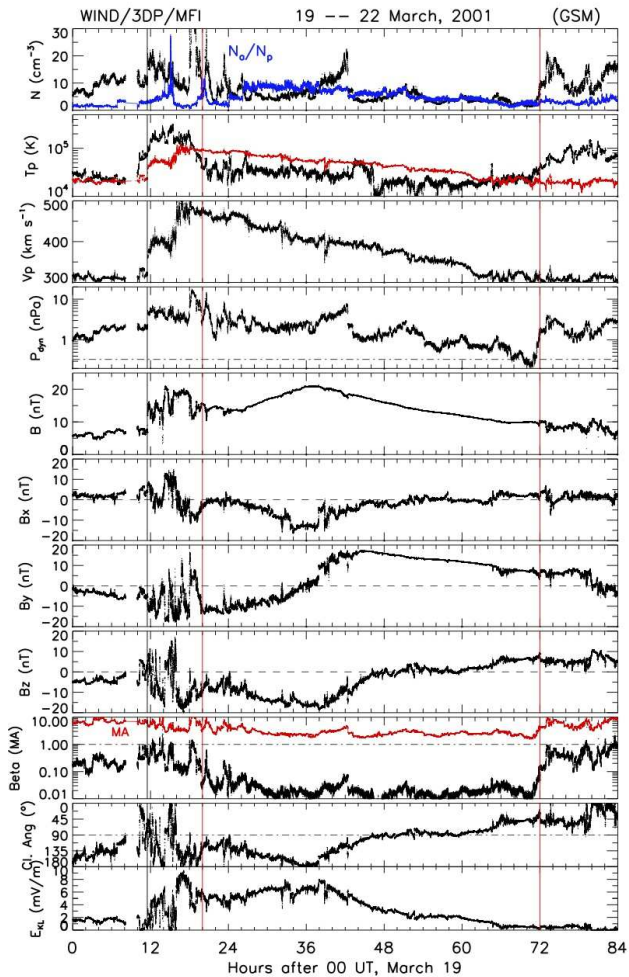}
 \caption{2001 March 19--22 long-duration event. From top to bottom, the panels show the proton density (alpha-to-proton ratio in blue), the proton temperature (expected temperature in red), the velocity, dynamic pressure, magnetic field strength and magnetic field vector components in GSM coordinates, the plasma $\beta$, magnetic field clock angle $\theta$ and the merging electric field.}
 \end{figure}
%%%%%%%%%%%%%

All but one lists of MCs consider this a single event with a duration of more than 50 hours \citep[see for example][]{Jian:2006, Marubashi:2007, Richardson:2008}, about twice longer than the typical duration of a magnetic cloud at 1~AU. The exception to this interpretation is the list of \citet{Lepping:2005}, which identifies two overlapping MCs (the second MC starts 0.5 hour before the end of the first one at 17:45UT on March 20). We performed a minimum-variance analysis on the magnetic field for the two separate intervals and found two inclined clouds with an angle of about $40^\circ$ between them with a large ratio of the intermediate to minimum eigenvalues. %Both MCs are left-handed and inclined, the first one more so than the second one (inclination of $-50^\circ$ and $-30^\circ$ respectively), and the second one was considered to last 45 hours.

This event resulted in a double-peaked intense geo-magnetic storm with Dst below $-50$~nT for 55~hours starting on March 19 at 18UT.  The first peak of $-105$~nT occurred  at 22UT on March 19 and the second peak of $-149$~nT at 14UT on March 20. The geomagnetic indices (AL and sym-H) are shown in Figure 4 as is the energetic particle flux at geostationary heights from GOES 8. The storm main phase starts when the sheath is passing over the magnetosphere. The sym-H index then decreases, though non-monotonically to its peak values. Recovery starts at the time when the second interval starts (after the maximum in $B$). In addition, this time period was associated with a sawtooth event on March 20 \citep[]{Troshichev:2011}. Sawtooth events are typically associated with a strong driver of the magnetosphere and a southward IMF for extended periods of time \citep[]{Henderson:2004}. There were 10 dipolarizations associated with injection of energetic particles observed at GOES 8 during the passage of the sheath and during the first interval. The average and standard deviation of the duration between individual sawteeth is 2.1 $\pm 0.55$~hours. During the second time interval, there are only four weak dipolarization events, none of them in the last 36 hours of the event.

%%%%%%
 \begin{figure}[t]
\centering
\includegraphics[width=7.2cm, height = 6cm]{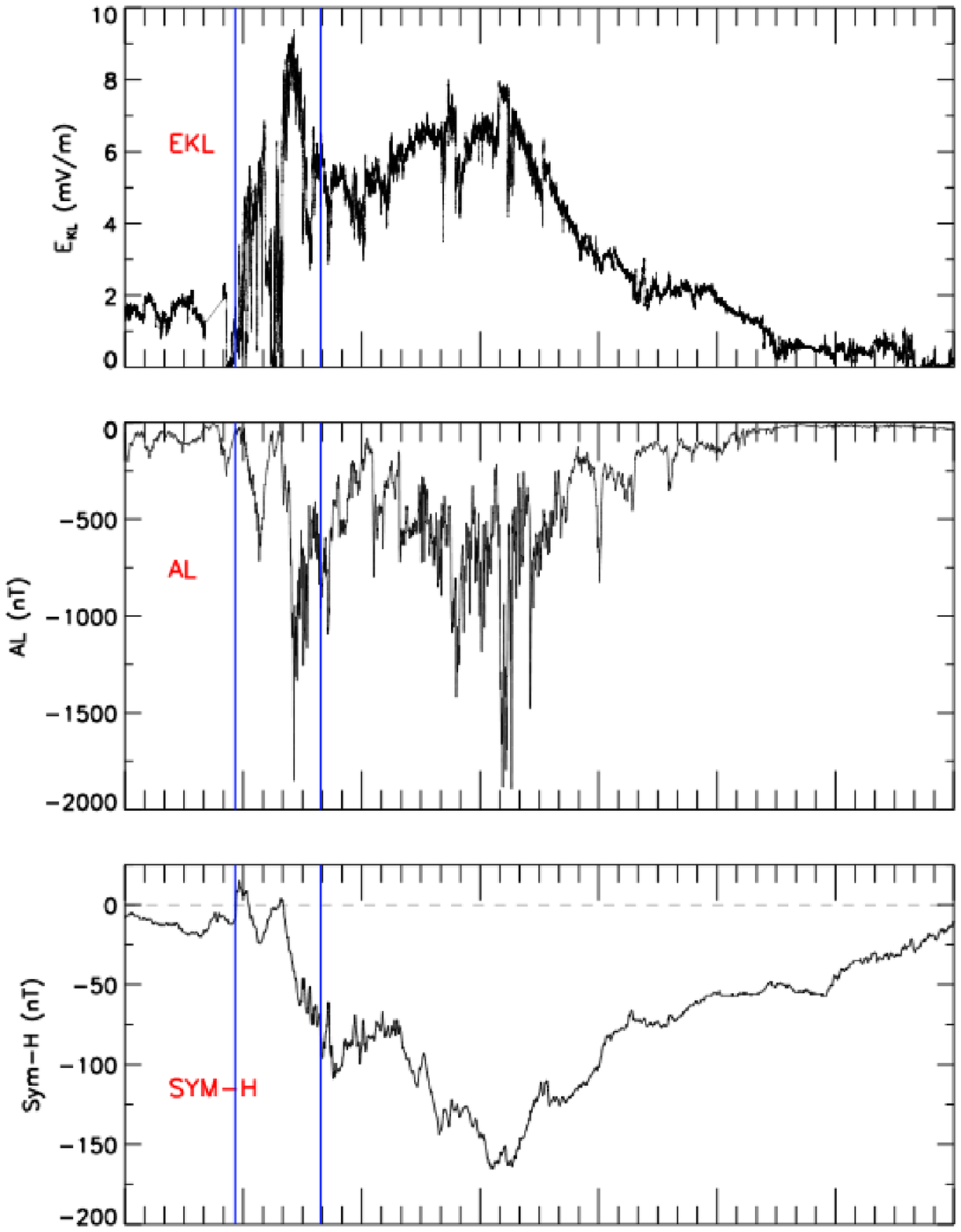}\\
\hspace{0.22cm}\includegraphics[width=7.3cm]{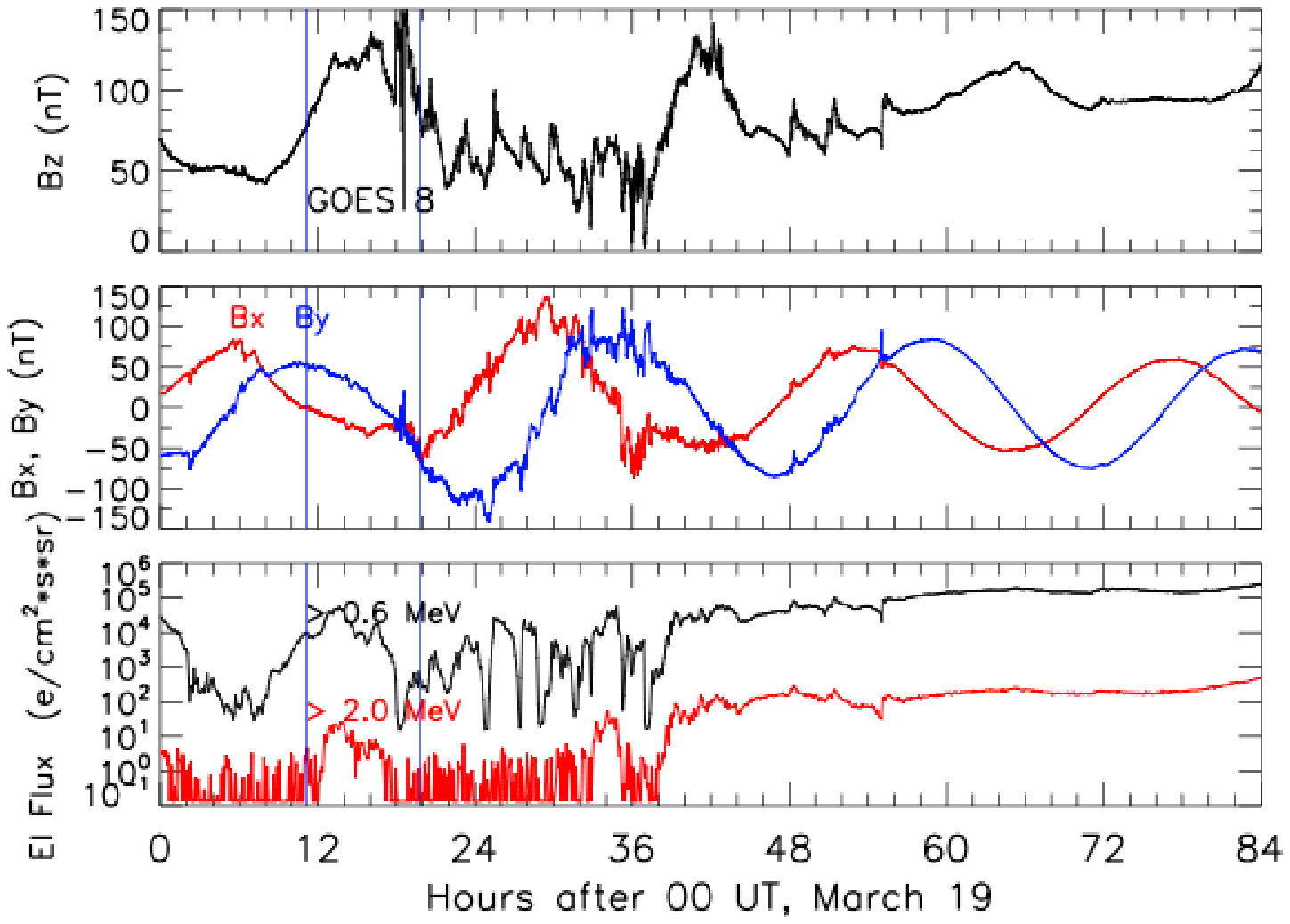}
 \caption{Geomagnetic response to the 2001 March 19--22 event. From top to bottom, the panels show the merging electric field, the AL and sym-H indices, the magnetic field components and energetic particle fluxes measured by GOES-8.}
 \end{figure}
%%%%%%%%%%%%%

The ejecta measured {\it in situ} in 2001 March 19--22 is usually associated with a slow partial-halo CME observed by LASCO/C2 on March 16 at 03:50UT with a speed of about 360~km~s$^{-1}$. While the transit time to 1~AU for this CME is approximately correct, it is unlikely that such a slow CME would (i) reach speeds in excess of 500~km~s$^{-1}$ at 1~AU, as is measured here, and (ii) result in the longest MC measured during solar cycle 23 \citep[]{Marubashi:2007}.
At the Sun, the period 2001 March 14--18 was CME-rich; there were a number of slow disk-centered eruptions on March 14--15 (for example from W10 on March 15 at 22:26UT) as well as a full and fast halo CME on March 18 at 02:00 UT. This halo CME lacks on-disk observations and is considered back-sided, but it may have been Earth-directed and could explain the origin of the second ejecta. Another, more plausible possibility is that the second ejecta corresponds to a CME from W37 on March 17 at 18:00UT which had a speed of about 600~km~s$^{-1}$.

Overall, there are many indirect indications that this event is associated with two CMEs, contrary to what was reported by most studies. It is the case although the velocity, magnetic field strength, proton temperature and plasma $\beta$ show no indication of two events. Based on the magnetic field component and the increase in proton density between 15 and 19UT on March 20, we can identify a first magnetic ejecta between 20UT on March 19 and 14UT on March 20 (18 hours) and a second ejecta between 19UT on March 20 and 00UT on March 22 (29 hours). As in the simulation, the second ejecta is characterized by a smooth, enhanced and uni-directional magnetic field (for the March event in the eastward direction). In the simulation, it corresponds to the direction of the axial field of a second CME for which the poloidal field has reconnected away leaving a nearly uni-directional field. If this is the case for the March event, the second CME would be of low inclination. The lack of geo-effectiveness of this second ejecta with a strong $B_y$ component is to further investigate.

\section{Discussion and Conclusions}\label{conclusion}

By combining numerical simulations and the analysis of {\it in situ} measurements, we have identified a new class of complex ejecta resulting from the interaction of two CMEs. Due to different orientations of the ejections, measurements at 1~AU appear to indicate the passage of a long magnetic cloud, but are in fact, due to two successive and interacting CMEs. With an appropriate orientation of the two CMEs, such an event may result in long-duration geomagnetic storm and be associated with sawtooth event.

In details, we have presented the results at 1~AU of a simulation of the interaction of two CMEs with different orientations. We have shown that the resulting complex ejecta is very similar to a MC from an isolated CME, except for the presence of a long ``tail'' in the magnetic field and the hotter temperature throughout the ejecta. We have estimated the expected Dst index for this complex ejecta, and we have found that, while the peak Dst is not as low as that from a well-oriented isolated CME, the tail in the magnetic field results in the Dst to be below $-100$~nT for more than a day, or about 50\% longer than for the isolated CME. 

We have also presented the analysis of one long-duration magnetic ejecta observed at 1~AU in 2001 March 19--22. This event resulted in a long, intense geomagnetic storm with a peak Dst of only $-149$~nT but the Dst stayed below $-50$~nT for more than 2 days. There were also a number of sawteeth in March 20 in the first half of the ejecta. Most studies have identified this as an isolated magnetic cloud with a duration in excess of 2 days. We have presented some potential evidence that this ejecta is in fact a complex ejecta associated with two CMEs. 

A more complete investigation of combined {\it in situ} and remote-sensing database will be required to assess how common this type of complex ejecta is. This could be helped in the future by the availability of remote-sensing observations of CMEs as they propagate and interact on their way to Earth \citep[]{CShen:2012, Lugaz:2012b}. The other event that we have tentatively identified in 2004 April 3--6 was also associated with an extended sawtooth event, although the Dst index peaked only at $-117$~nT and was below $-50$~nT for only 15 hours. Further studies are also required to determine how this type of complex ejecta affects Earth's magnetosphere and how the interaction differs from that with an isolated MC or a multiple-MC event. 

%%% End of body of article:

%%%%%%%%%%%%%%%%%%%%%%%%%%%%%%%%
%% Optional Appendix goes here
%
% \appendix resets counters and redefines section heads
% but doesn't print anything.
% After typing  \appendix
%
% \section{Here Is Appendix Title}
% will show
% Appendix A: Here Is Appendix Title
%

%  ACKNOWLEDGMENTS

\begin{acknowledgments}
The research for this manuscript was supported by the following grants: NSF AGS-1239699 and NASA NNX13AH94G.
The simulations were performed on the NASA HEC {\it Pleiades} system under awards SMD-12-3360 and 13-3919. 
\end{acknowledgments}

%% ------------------------------------------------------------------------ %%
%%  REFERENCE LIST AND TEXT CITATIONS
%

\bibliographystyle{agufull08}
%\bibliography{thesis}

%% ------------------------------------------------------------------------ %%
%
%  END ARTICLE
%
%% ------------------------------------------------------------------------ %%
\end{article}

\end{document}